\documentclass[11pt]{article}
\usepackage{graphicx}
\usepackage{amssymb}
\usepackage{amsmath}
\usepackage{epsfig}

\begin{document}
\title{Accelerating cosmologies tested by distance measures}


\author{V. Barger$^{1}$, Y. Gao$^{1}$ and D. Marfatia$^{2}$\\[2ex]
\small\it $^1$Department of Physics, University of Wisconsin, Madison, WI 53706\\
\small\it $^2$Department of Physics and Astronomy, University of Kansas, Lawrence, KS 66045}

\date{}

\maketitle

\begin{abstract}
We test if the latest Gold set of 182 SNIa or the combined ``Platinum'' 
set of 192 SNIa from the ESSENCE and Gold sets, in conjunction with the CMB shift parameter show a preference 
between the $\Lambda$CDM model,
three $w$CDM models, and the DGP model of modified gravity as an explanation for the current accelerating phase of 
the universe's 
expansion. We consider flat $w$CDM models with an equation of state $w(a)$ that is (i) constant with scale factor $a$, 
(ii) varies as $w_0+w_a(1-a)$ for redshifts probed by supernovae but is fixed at $-1$ at earlier epochs and (iii)
varies as $w_0+w_a(1-a)$ since recombination. We find that all five models explain the data with
comparable success. 

\end{abstract}
\newpage

Observations of Type Ia supernovae (SNIa), the cosmic microwave background, and large scale structure corroborate that the expansion 
of our universe is accelerating. 
Explanations of this acceleration often invoke the existence of a dark energy component
with the unusual property of negative pressure~\cite{Peebles:2002gy}. 
The simplest example of dark energy is the cosmological constant $\Lambda$ with an 
equation of state $w\equiv p/\rho=-1$. Although in the context of $\Lambda$, $p$ and $\rho$ are the effective pressure and 
energy density attributed to vacuum energy, in general they are the corresponding quantities of the fluid describing dark energy. 
Possibilities
with a time-varying equation of state are modelled by introducing a scalar field called quintessence~\cite{quint}. 
No dark energy model has a natural explanation. 
Explanations of the current acceleration that avoid the introduction of dark energy emerge from modifications to gravity, typically
in the context of braneworld models.   

For models with dark energy, we consider the nonflat $\Lambda$CDM model, and three flat universe 
models with equations of state that differ from $-1$ at some time in the expansion history. We refer to these as flat $w$CDM models. 
The simplest of these has a constant $w$ that differs from $-1$. To model dark energy with a time-varying $w$, 
we adopt a 
parameterization that is well-behaved at high redshift: $w(a)=w_0+w_a(1-a)$~\cite{param}, 
where $a$ is normalized such that it
is equal to $1$ today and is related to redshift $z$ by $a=1/(1+z)$. 
While this parameterization reduces to the
linear relation $w(z)=w_0+w_a z$~\cite{oldparam} at low redshifts, it has the disadvantage that when analyzing
high-redshift data,
we are implicitly restricting ourselves to models that are well-represented by this time evolution. Said differently, we
are placing the strict prior that dark energy must have evolved in this manner throughout. 
Keeping the latter in mind, one case 
we consider is that dark energy evolves according to $w(a)=w_0+w_a(1-a)$ for redshifts probed by SNIa and behaves like a 
cosmological constant at earlier
epochs. A second case, involving a stronger prior, is that $w(a)=w_0+w_a(1-a)$ is valid till the epoch of last scattering.

The Friedman equation in dimensionless form with $H\equiv \dot{a}/a$ and $H_0\equiv H(z=0)$ is
\begin{eqnarray}
\frac{H(z)^2}{H_0^2}&=&\Omega_m(1+z)^3+\Omega_w(1+z)^{3(1+w_0+w_a)}e^{-3w_a {z\over{1+z}}}
+\Omega_k(1+z)^2\,,
\end{eqnarray}
where $\Omega_m$, $\Omega_w$ and $\Omega_k$ are the matter, dark energy and curvature 
densities in units of the critical density.
For the $\Lambda$CDM model, $\Omega_w \equiv \Omega_\Lambda$, $w_0=-1$ and $w_a=0$. For flat models, 
\mbox{$\Omega_k=1-\Omega_m-\Omega_w=0$}.

The DGP model~\cite{Dvali:2000hr} (named so for its authors) is a generally-covariant infrared modification of 
general relativity (GR) and 
is not reducible to an extension of GR with additional scalar or vector degrees of freedom. 
At short distances gravity is 
4-dimensional, while in the far infrared, gravity appears to
be 5-dimensional because of gravitational leakage from the 4-dimensional brane into the 5-dimensional bulk. 
The implications for cosmology are that at early times the
correction from the infrared modification is negligible and the universe obeys the standard cosmology. However, 
at late times, the weakening of gravity is significant and leads to self-accelerated expansion without the need for any form of 
matter~\cite{Deffayet:2000uy}. Due its firm theoretical foundation~\cite{Lue:2005ya}, we analyze the DGP model as the 
canonical example of modified
GR that explains the current acceleration without the need for dark energy.

The Hubble expansion is given by~\cite{Deffayet:2000uy}
\begin{eqnarray}
\frac{H(z)^2}{H_0^2}&=&[\sqrt{\Omega_r}+\sqrt{\Omega_r+\Omega_m(1+z)^3}]^2+\Omega_k(1+z)^2\,.
\label{dgp}
\end{eqnarray}
Here $\Omega_r\equiv (4r_c^2H_0^2)^{-1}$, where $r_c\equiv M_4^2/(2M_5^3)$ is the length scale beyond which 4-dimensional gravity
(with Planck scale $M_4$) transits to 5-dimensional gravity (with Planck scale $M_5$). Setting $z=0$ in Eq.~(\ref{dgp})
 yields \mbox{$\Omega_k= 1-[\sqrt{\Omega_r}+\sqrt{\Omega_r+\Omega_m}]^2$.} Note that the DGP model has the same number of parameters
as $\Lambda$CDM, with $\Omega_r$ replacing $\Omega_\Lambda$. 
 
To compare the various models on an equal-footing, we only analyze data that probe the expansion history of the universe. We do
not utilize data that are sensitive to the evolution of density perturbations since these have not been fully 
worked out for the DGP model, 
although progress has been made in Ref.~\cite{Song:2006jk}.
We analyze the distance moduli of the Gold set of 182 SN~\cite{gold} (of which 16 have $z>1$) 
compiled from Refs.~\cite{Riess:2006fw,Astier:2005qq,Riess:2004nr} and the ``Platinum'' set of
192 SN~\cite{platinum} compiled from the ESSENCE~\cite{essence} and Gold sets.
In our analyses, we include the CMB shift parameter~\cite{Bond:1997wr} which 
measures the distance to the last scattering surface. For most of what follows, 
we do not use the baryon acoustic oscillation (BAO) 
distance parameter $A$ extracted
from the scale corresponding to the first acoustic peak at recombination~\cite{Eisenstein:2005su}. 
The procedure used to determine $A$ assumes
that $w$ does not vary with redshift. The resulting value may not be applicable for time-varying $w$~\cite{Dick:2006ev}, 
making it unsuitable for three of the five models we are considering. 
Recent joint analyses of the DGP model with older data 
in various combinations (with some including the $A$ parameter) can be found in Ref.~\cite{older}.

The shift parameter, defined in terms of the $H_0$-independent luminosity distance $D_L= H_0 d_L$ 
(where $d_L$ is the luminosity distance)~\cite{Bond:1997wr}: 
\begin{equation}
R\equiv \sqrt{\Omega_m} \frac{D_L(z_{CMB})}{(1+z_{CMB})}\,,
\end{equation}
is approximately equivalent to the ratio of the sound horizon at recombination to the comoving distance to the last scattering
surface. For all practical purposes, $R$ is model-independent. We use the value $R=1.70\pm 0.03$~\cite{Wang:2006ts} obtained 
from the WMAP 3-year data~\cite{Spergel:2006hy} with a redshift at recombination $z_{CMB}=1089\pm 1$. 

The statistical significance of a model is determined by evaluating $\chi^2_R=(R^{obs}-R^{th})^2/\sigma_R^2$ and $\chi^2_{SN}$, 
which after marginalization over a nuisance parameter, has the absolute value~\cite{DiPietro:2002cz}
\begin{equation}
\chi^2_{SN}=A-\frac{B^2}{C}\,,
\end{equation}
where
\begin{eqnarray}
A=\sum_{i=1}^{N} {\frac{(\mu_i^{obs}-5\log_{10}D_L(z_i))^2}{\sigma_i^2}}\,,\ \ \ \
B=\sum_{i=1}^{N} {\frac{\mu_i^{obs}-5\log_{10} D_L(z_i)}{\sigma_i^2}}\,,\ \ \ \
C=\sum_{i=1}^{N} {\frac{1}{\sigma^2_i}}\,\nonumber.
\end{eqnarray}
Here, $\mu_i^{obs}$ and $\sigma_i$ are the distance modulus and its uncertainty at redshift $z_i$. 
$N=182$ for the Gold set and $N=192$ for the Platinum set.
\\

\begin{table}[t]
\begin{center}
\begin{tabular}{|l|c c c| c c c|}
\hline 
              & Gold   & (182 SN)&  & Platinum &(192 SN) & \\ \hline
\hline
 $\Lambda$CDM & $\chi^2$ & $\Omega_{\Lambda}$ & $\Omega_m$ & $\chi^2$ & $\Omega_{\Lambda}$ & $\Omega_m$  \\ \hline
 SN           & 156.4 & 0.95 & 0.48 & 195.2 & 0.85 & 0.33 \\
 SN+$R$       & 158.4 & 0.68 & 0.36 & 195.6 & 0.74 & 0.27 \\ 
SN+$R$+BAO    & 161.4 & 0.72 & 0.30 & 195.6 & 0.74 & 0.27 \\ \hline
 $w$CDM, $w_a=0$ & $\chi^2$ & $w_0$ & $\Omega_m$ & $\chi^2$ & $w_0$ & $\Omega_m$\\ \hline
 SN              & 156.6 &  -1.75 & 0.46 & 195.4 & -1.16 & 0.31 \\
 SN+$R$          & 160.2 & -0.85 & 0.28 & 195.9 & -0.94 & 0.24 \\
SN+$R$+BAO    & 160.3 & -0.86 & 0.29 & 196.6 & -0.98 & 0.26 \\ \hline
 $w$CDM, $w(z>1.8)=-1$ & $\chi^2$ & $w_0$ & $w_a$  & $\chi^2$ & $w_0$ & $w_a$\\ \hline
 SN              & 156.5 & -1.11 & 2.39 & 195.3 & -1.11 & -1.16 \\
 SN+$R$          & 156.5 & -1.28 & 2.69 & 195.5 & -1.06 & 0.81 \\ \hline
 $w$CDM          &  $\chi^2$ & $w_0$ & $w_a$ &  $\chi^2$ & $w_0$ & $w_a$  \\ \hline
 SN              & 156.5 & -1.11 & 2.39 & 195.3 & -1.11 & -1.16 \\
 SN+$R$          & 157.0 & -1.35 & 1.54 & 195.5 & -1.09 & 0.67 \\ \hline
 DGP             & $\chi^2$ & $\Omega_r$ & $\Omega_m$& $\chi^2$ & $\Omega_r$ & $\Omega_m$  \\ \hline
 SN              & 156.4 & 0.24 & 0.36 & 195.1 & 0.22 & 0.24 \\
 SN+$R$          & 160.3 & 0.14 & 0.23& 196.4 & 0.16 & 0.17 \\ \hline
\end{tabular}
\label{tab:chi2}
\end{center}
\caption{The minimum $\chi^2$ values and best-fit parameters for each of the five models. The $\chi^2$ per degree
of freedom is substantially different for the Gold and Platinum datasets. All $w$CDM models 
are flat. In the analyses
in which both $w_0$ and $w_a$ are allowed to vary freely, we require that $\Omega_m$  take values 
between 0.15 and 0.35. All five models have $\chi^2_R=0$ at the minimum in the analysis of the shift parameter alone. 
We do not show the corresponding best-fit parameters
because the $\chi^2_R$ distributions are too broad for the parameters to be meaningful. We have included the BAO 
constraint for the two models with a constant $w$ for comparison. The number of parameters
in the analyses of the successive models are $3,3,4,4,3$, respectively, 
where in each case one parameter
is a nuisance parameter that is marginalized over.}
\end{table}

{\bf{$\Lambda$CDM:}} In Fig.~\ref{fig:lcdm} we display the results of our analysis of the nonflat $\Lambda$CDM model. 
The shaded
regions are the $1\sigma$, $2\sigma$ and $3\sigma$ allowed regions from an analysis of the latest SN data 
(upper panel: Gold set, lower panel: Platinum set) 
and of the CMB shift parameter $R$. Although expected, the orthogonality of the regions is striking. 
The solid contours depict the corresponding regions from the joint analysis.
The best-fit parameters are provided in Table~1. 
The dot marks the best-fit from the joint analysis.
Notice that the combined analyses prefer universes that are almost flat with a tendency for positive curvature. 
If we restrict
ourselves to flat universes, the minimum $\chi^2$ value for the joint analysis with the Gold set (Platinum set) 
increases to 163 (196.3), 
and the best-fit moves to $\Omega_m=0.3$ ($\Omega_m=0.25$); $\chi^2_{SN}$ increases to 158.6 for the Gold set but remains
essentially unchanged
for the Platinum set. 

{\bf{Flat $w$CDM with constant $w$:}} We allow $w$ to take values different from $-1$, but do not allow for time variation by 
setting $w_a=0$. Figure~\ref{fig:wcdm1} is similar to Fig.~\ref{fig:lcdm} except that we plot $w_0$ vs $\Omega_m$.
Keeping in mind that the number of degrees of freedom is the same as that for the 
$\Lambda$CDM model, we see that the fit to the $\Lambda$CDM model is not significantly better; with the Gold set 
(Platinum set) it has a minimum 
$\chi^2$ that is 
only 1.8 (0.3) lower.

{\bf{Flat $w$CDM with late-varying $w$:}} In this case, $w$ has time-variation only from $z=1.8$ until today, and equals $-1$
at earlier times. We also require $0.15\le \Omega_m \le 0.35$. In Fig.~\ref{fig:wcdm2}, we only show 
the $1\sigma$ and $2\sigma$ regions
for the separate SN and $R$ analyses for obvious reasons. 
 It is no surprise that the $R$ parameter shows no sensitivity to
$w_a$ because we have assumed that the dark energy behaves as a cosmological constant since 
recombination until $z=1.8$. This 
assumption eliminates the constraining power of $R$ that comes from its long lever arm. The joint constraint is dominated by the
SN data. This is evident from Fig.~\ref{fig:wcdm2} and Table~1. The allowed regions and best-fit point do not move 
significantly on adding the 
$R$ parameter to the analysis. Although the Gold and Platinum sets favor different regions of $w_a$, 
on comparing the allowed regions with the regions occupied by different dark energy 
models in the ($w_0,w_a$) 
plane, as classified in Ref.~\cite{Barger:2005sb}, it is clear that no class of models is excluded even at the 
$2\sigma$ C.~L. by either dataset.

{\bf{Flat $w$CDM with varying $w$:}} The equation of state is allowed to vary since recombination. Again, we 
require $0.15\le \Omega_m \le 0.35$. 
As expected,  the results of analyzing SN data alone are identical with that of late-varying $w$ since there are no 
SN data with $z>1.8$. See Table~1 and Figs.~\ref{fig:wcdm2} and~\ref{fig:wcdm3}. 
From Fig.~\ref{fig:wcdm3}, we now see how $R$ helps to constrain $w_a$. Nevertheless, all the
dark energy models classified in Ref.~\cite{Barger:2005sb} remain safe. 
With respect to $\Lambda$CDM, the additional free 
parameter improves the minimum $\chi^2$ of the joint analysis with the Gold set (Platinum set) by only 1.4 (0.1).
 
{\bf{DGP:}} Figure~\ref{fig:dgp} is similar to Fig.~\ref{fig:lcdm} with $\Omega_\Lambda$ replaced by the 
physical parameter relevant to DGP, $\Omega_r$. A similar orthogonal relationship exists between the SN and $R$ regions.
However, here we see that the joint analysis with the Gold set (Platinum set) prefers a slightly open universe with 
$\Omega_k=0.027\pm 0.014$ ($\Omega_k=0.041\pm 0.010$). 
The overall fit is only slightly worse compared to $\Lambda$CDM. 
\\

To assess the impact of the BAO constraint on the two models that do not have a time-varying equation of state 
($\Lambda$CDM and the $w$CDM model with constant $w$), we use the measured
value $A=0.469\pm 0.017$~\cite{Eisenstein:2005su}, where
\begin{equation}
A\equiv \sqrt{\Omega_m} \bigg[\frac{H_0}{H(z_{BAO})} \bigg(\frac{D_L(z_{BAO})}{z_{BAO}(1+z_{BAO})}\bigg)^2\bigg]^{1/3}\,,
\end{equation}
to calculate $\chi^2_A=(A^{obs}-A^{th})^2/\sigma_A^2$. Here, $z_{BAO}=0.35$ is the typical redshift of the SDSS sample
of luminuous red galaxies. From Figs.~\ref{fig:bao1} and~\ref{fig:bao2}, it can be seen that the BAO constraint
provides a satisfying confirmation that the different datasets are concordant, and helps to further constrain 
 the regions from the joint analysis of SN data and $R$. The minimum $\chi^2$ and best-fit parameters from the joint
analyses including the BAO constraint are provided in Table~1. For the $\Lambda$CDM model, 
the addition of this datapoint to the joint analysis with the Gold set results in a preference for a universe 
with less curvature at the expense of increasing
the minimum $\chi^2$ by 3. The BAO constraint has no effect on the best-fit parameters of the $\Lambda$CDM model 
in a joint analysis with the Platinum set.

\vskip 0.2in
In conclusion, it is noteworthy that all 5 models fit the SN data equally well. 
Only on inclusion of $R$ in the analyses do
minor differences develop. Current data cannot tell if the accelerated expansion is caused by a cosmological constant, by 
 dark energy with a constant $w$, by dark energy whose equation of state started varying recently or has always been varying, or due
to modified gravitational physics of conventional matter{\footnote{Other analyses of
similar datasets can be found in Ref.~\cite{catchall}.}. A detailed understanding of how density perturbations evolve in braneworld
cosmologies will enable
the use of the vast amount of data available on the power spectrum from observations of the CMB and large scale structure and may 
help with the basic question of whether the acceleration is due to a new form of energy or due to new aspects of gravity. 
A step forward from kinematical probes of modified gravity to dynamical ones is the order of the day.

\vskip 0.2in
{\bf{Acknowledgments.}}
We thank Ned Wright for a communication. 
This research was supported by the U.S.
Department of Energy under Grants No. DE-FG02-95ER40896 and
DE-FG02-04ER41308 and by the NSF under CAREER Grant No. PHY-0544278.
Computations were performed on facilities supported by the NSF under
Grants No. EIA-032078 (GLOW), PHY-0516857 (CMS Research Program
subcontract from UCLA), and PHY-0533280 (DISUN), and by the Wisconsin Alumni Research Foundation.


\newpage

\begin{figure}[hbp]
 \begin{center}
 \epsfysize=4truein
  \epsfxsize=5truein
    \epsffile{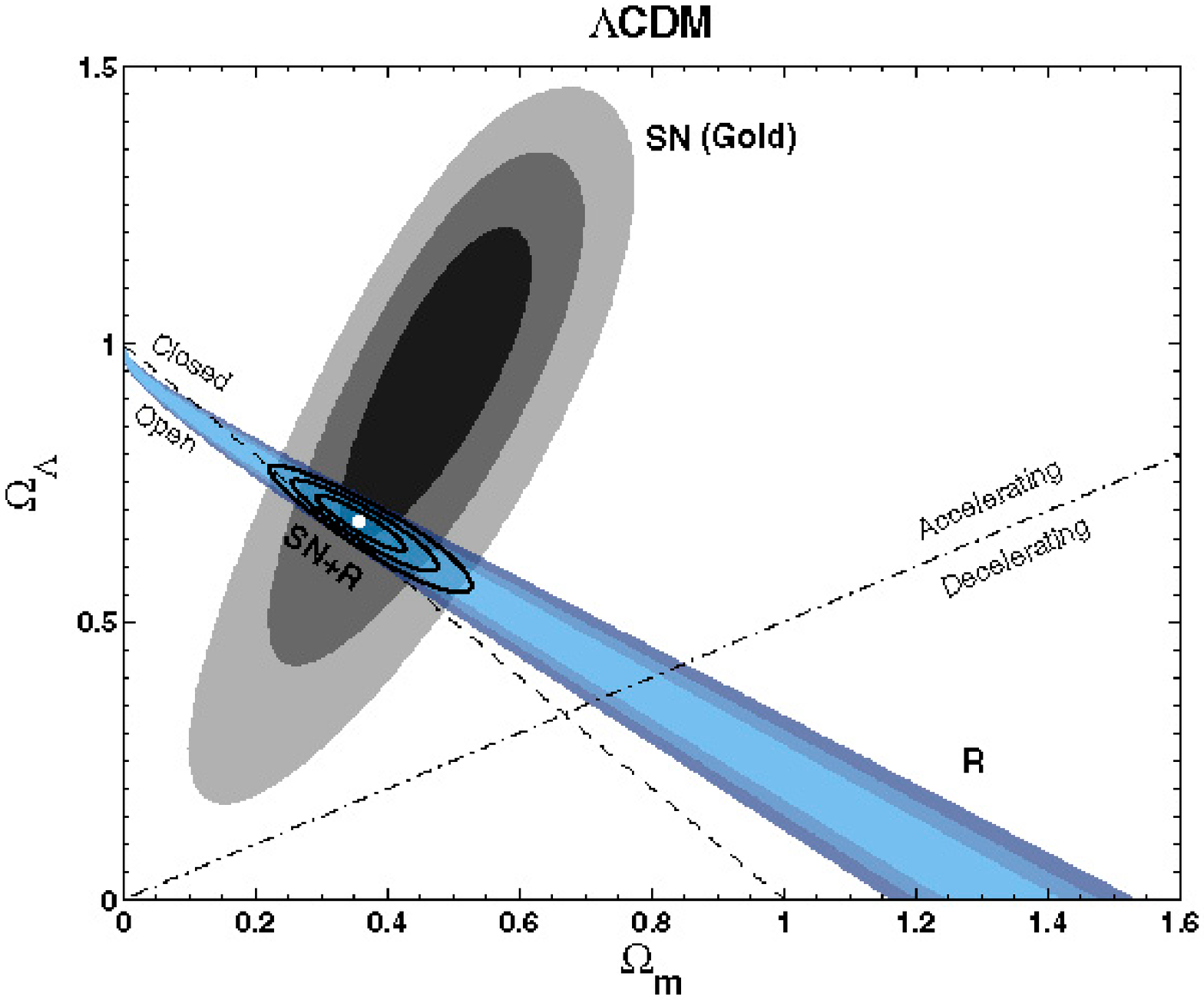}, 
    \epsfysize=4truein
  \epsfxsize=5truein 
   \epsffile{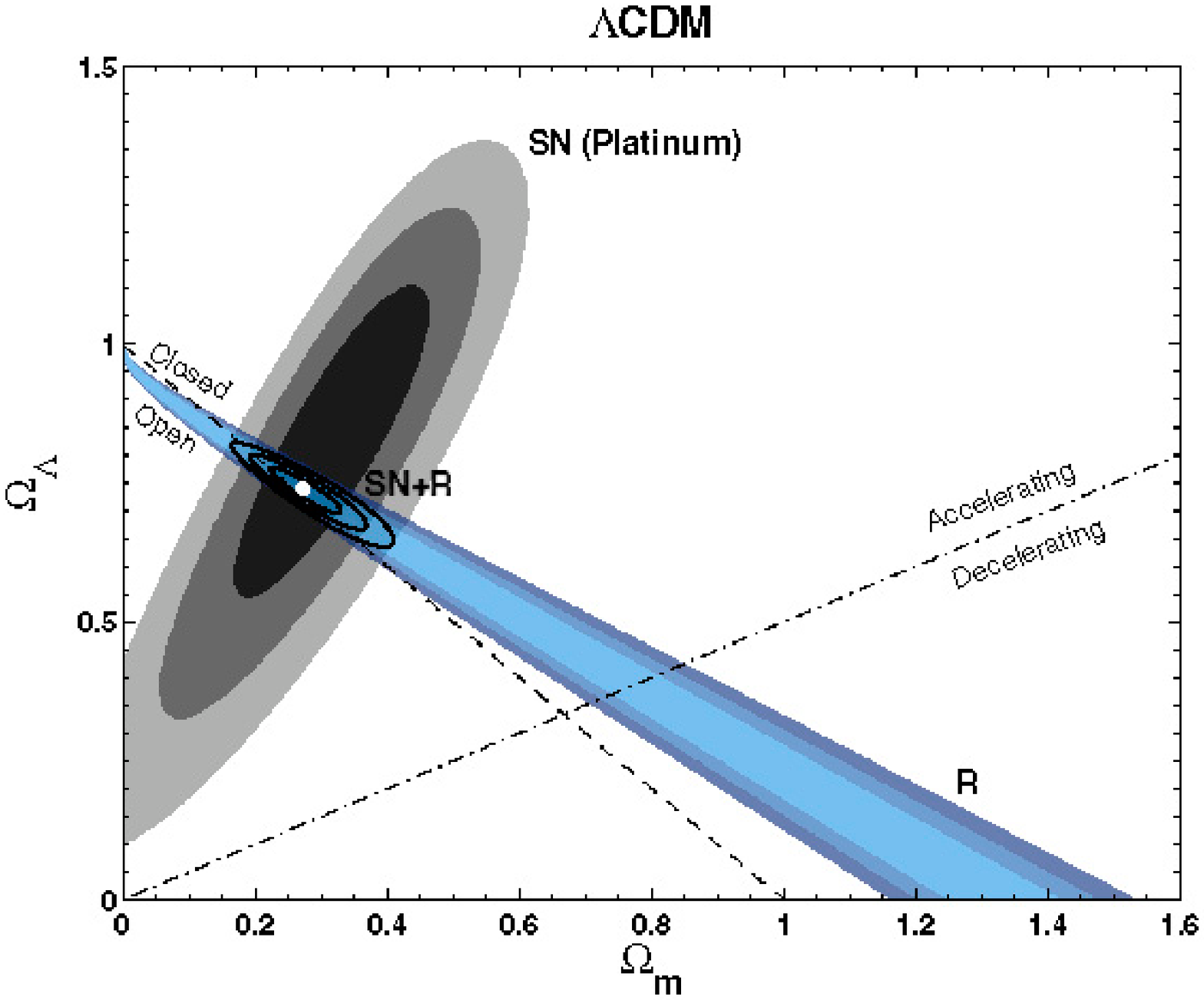}
\caption[]{The $1\sigma$, $2\sigma$ and $3\sigma$ allowed regions for the $\Lambda$CDM model from an analysis of 
SN data (upper panel: Gold 
set, lower panel: Platinum set) of the CMB shift parameter $R$ and from a joint analysis. 
The best-fit parameters from the joint analyses 
are indicated by a dot. See Table~1.
\label{fig:lcdm}}
\end{center}
\end{figure}

\begin{figure}[hbp]
 \begin{center}
  \epsfysize=4truein
  \epsfxsize=5truein
    \epsffile{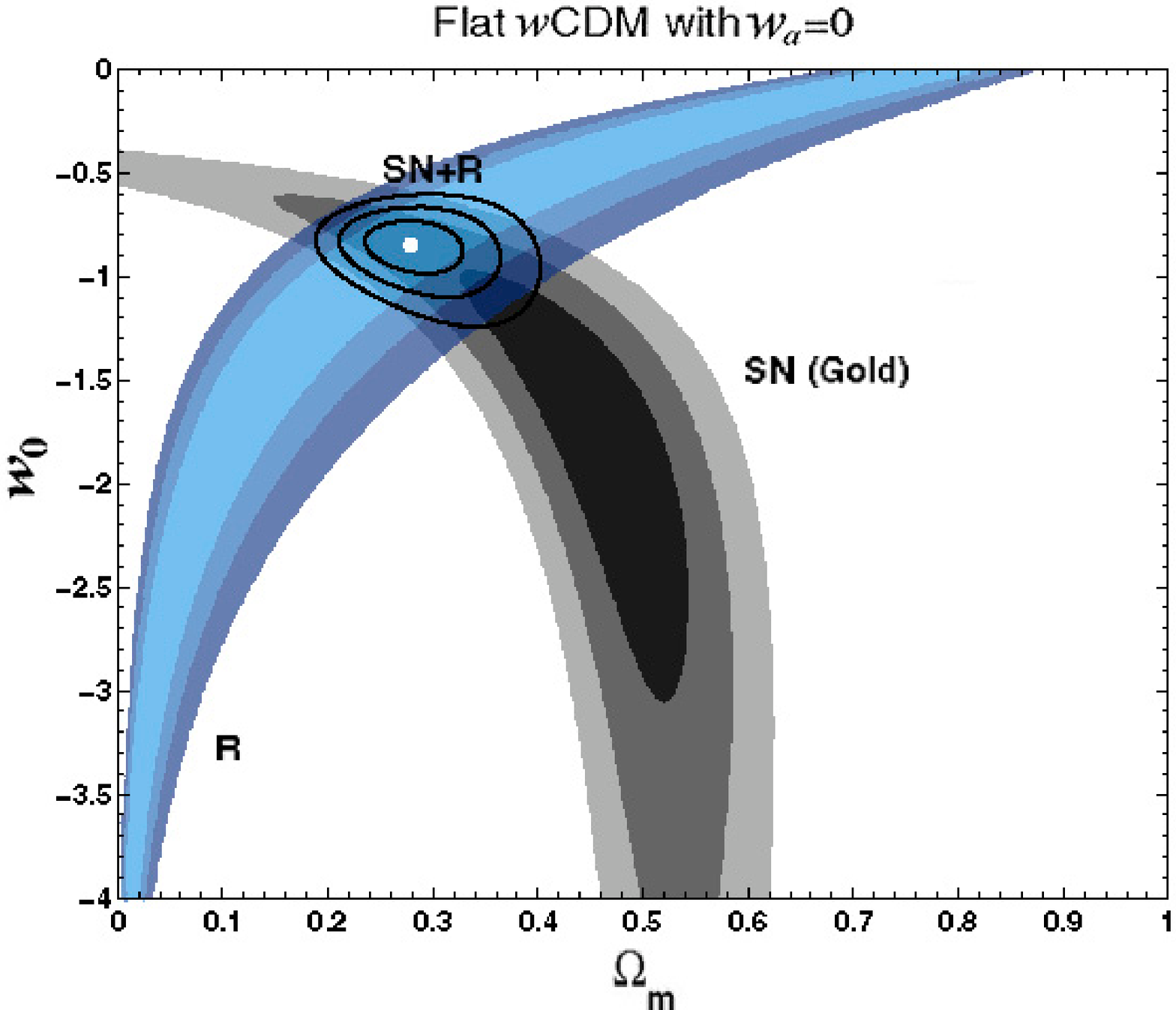},
  \epsfysize=4truein
  \epsfxsize=5truein
    \epsffile{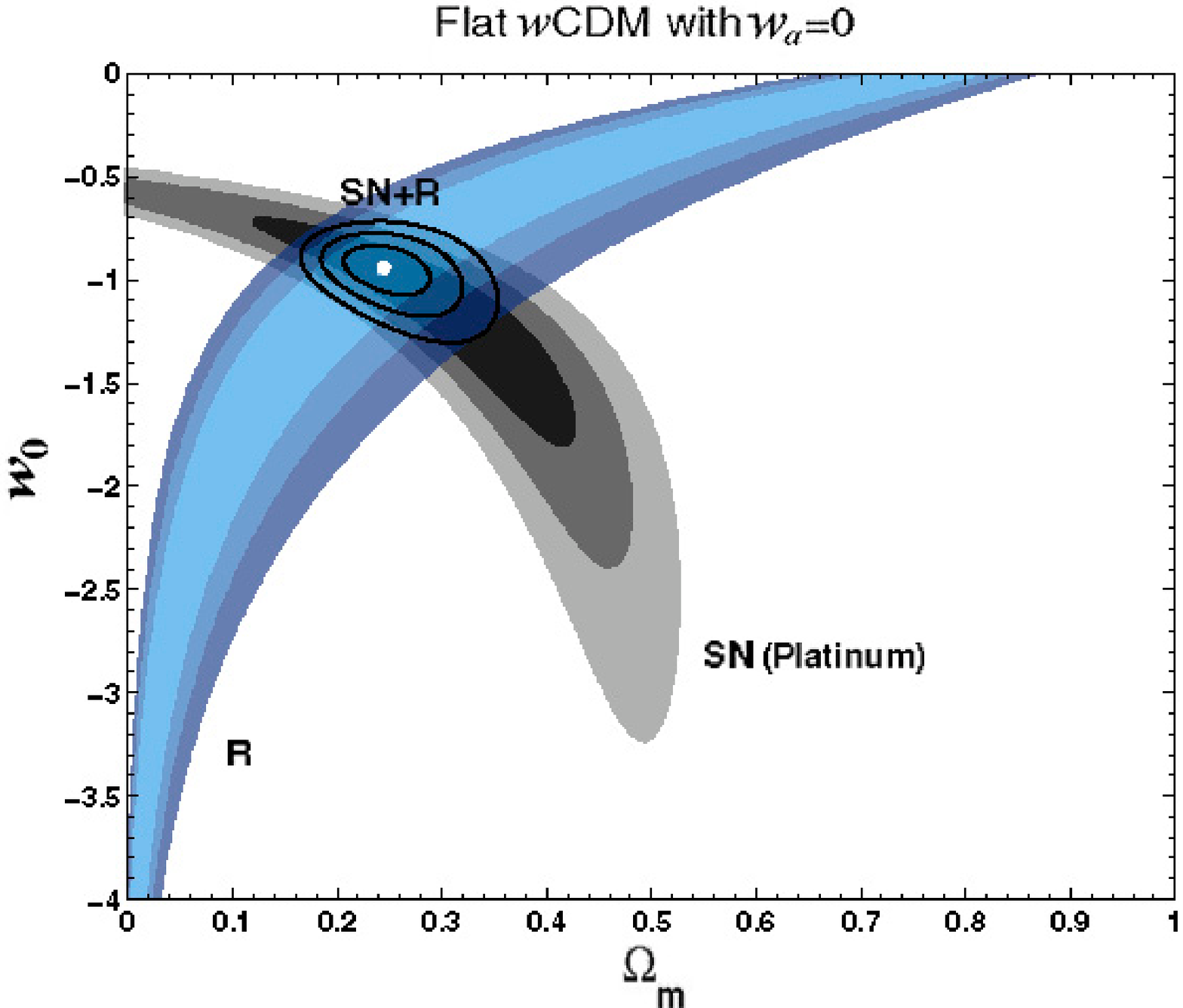}
\caption[]{Similar to Fig.~\ref{fig:lcdm}, but for the $w$CDM model with a constant equation of state $w(a)=w_0$.
\label{fig:wcdm1}}
\end{center}
\end{figure}

\begin{figure}[hbp]
 \begin{center}
  \epsfysize=4truein
  \epsfxsize=5truein
    \epsffile{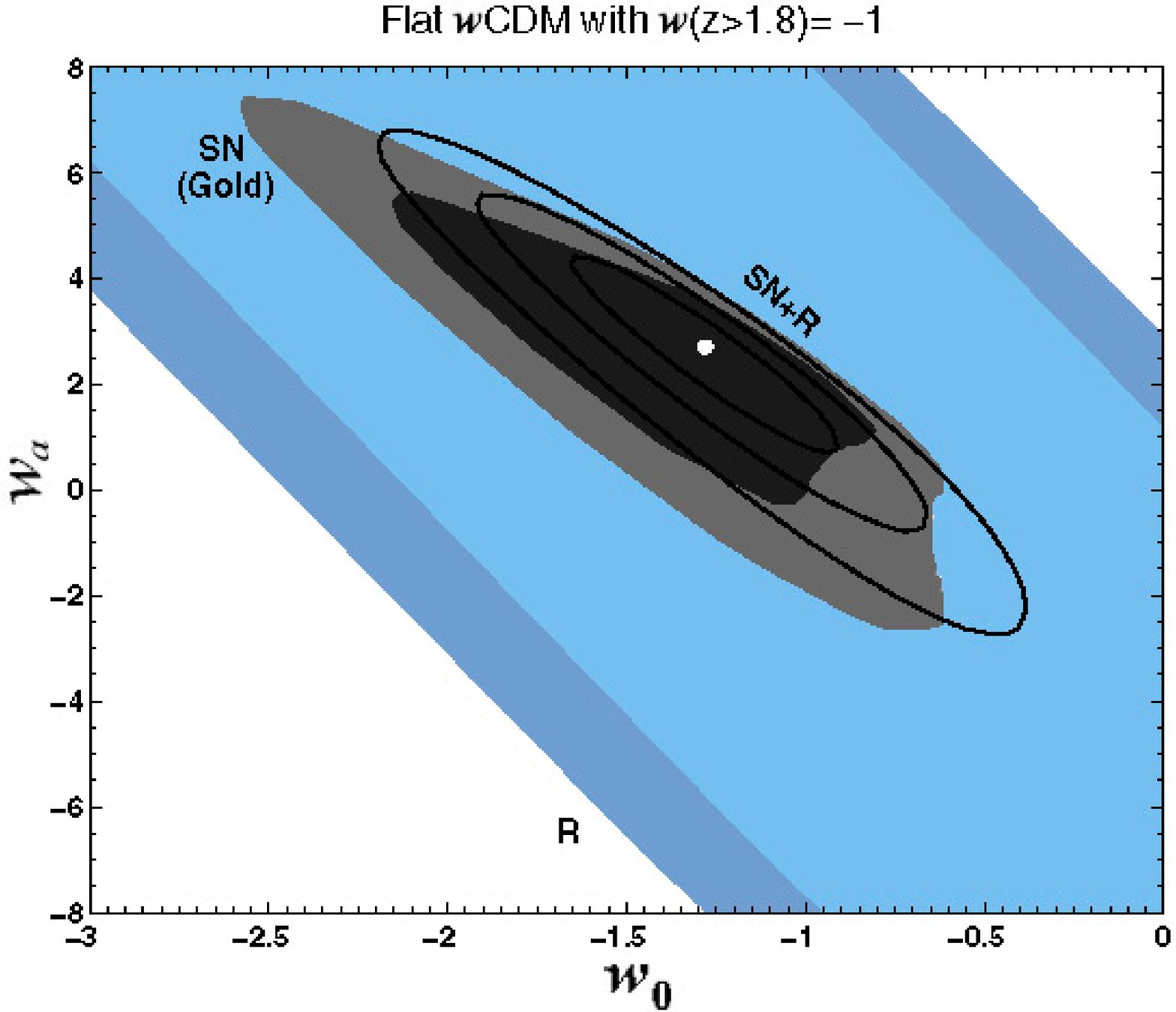},
  \epsfysize=4truein
  \epsfxsize=5truein
    \epsffile{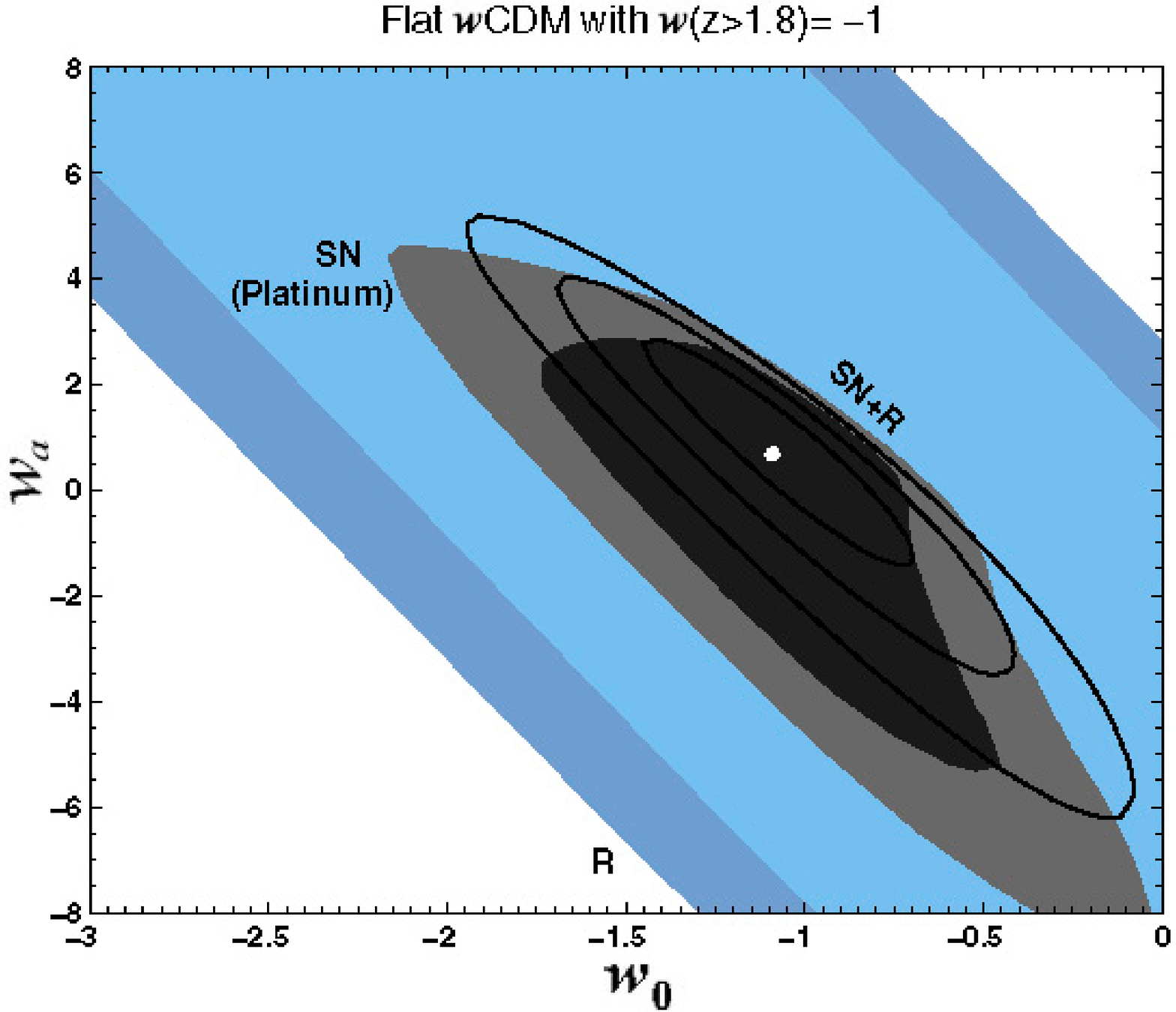}
\caption[]{The allowed regions for the flat $w$CDM model with an equation of state $w(a)=w_0+w_a(1-a)$ 
in the redshift range $0\le z \le 1.8$ and a constant equation of state $w=-1$ for $z>1.8$. Only the $1\sigma$ and $2\sigma$ regions
are shown for separate SN and $R$ analyses, while the $3\sigma$ region is also shown for the joint analysis. 
The best-fit parameters from the joint analysis 
are indicated by a dot.
\label{fig:wcdm2}}
\end{center}
\end{figure}

\begin{figure}[hbp]
 \begin{center}
  \epsfysize=4truein
  \epsfxsize=5truein
    \epsffile{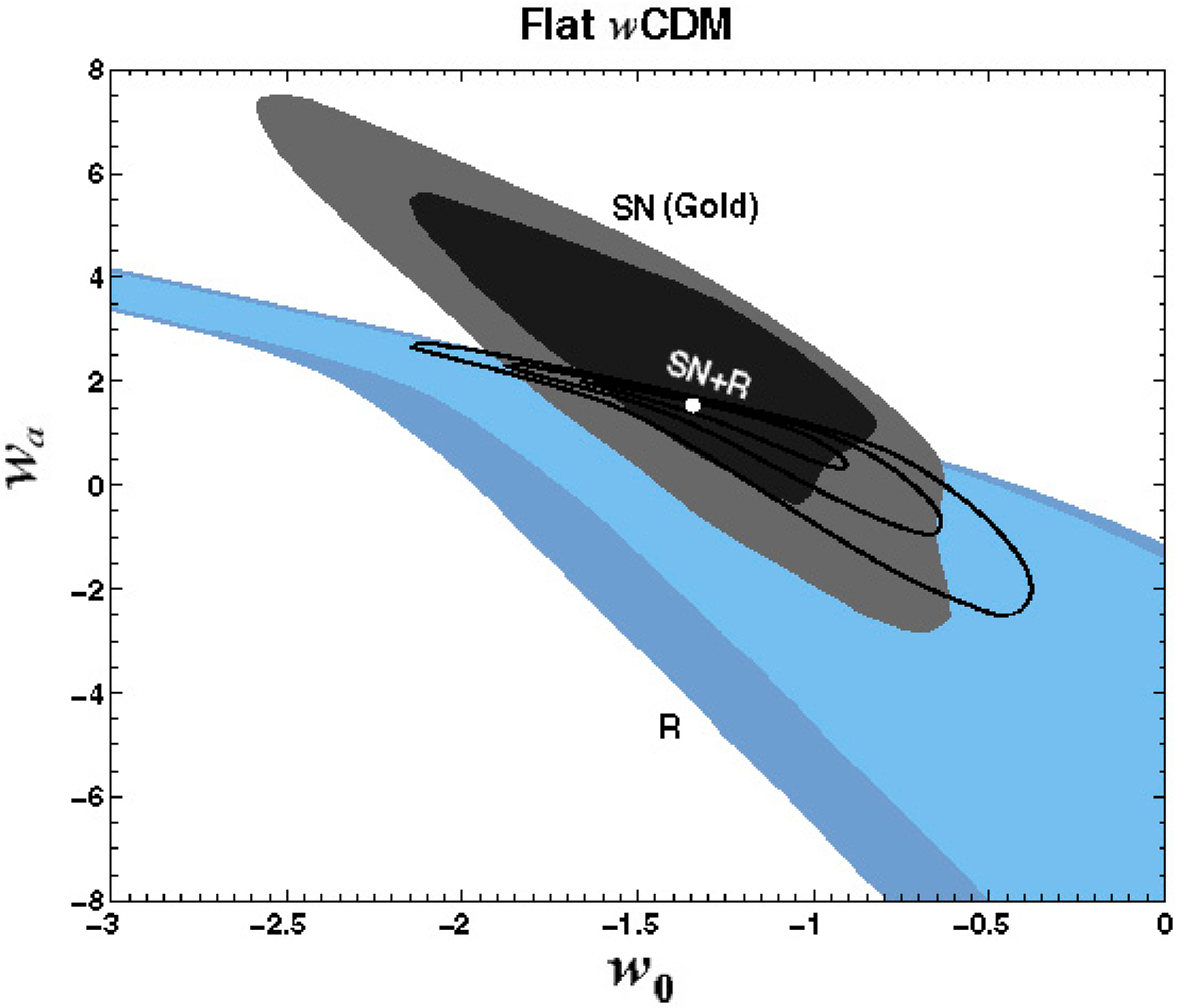},
  \epsfysize=4truein
  \epsfxsize=5truein
    \epsffile{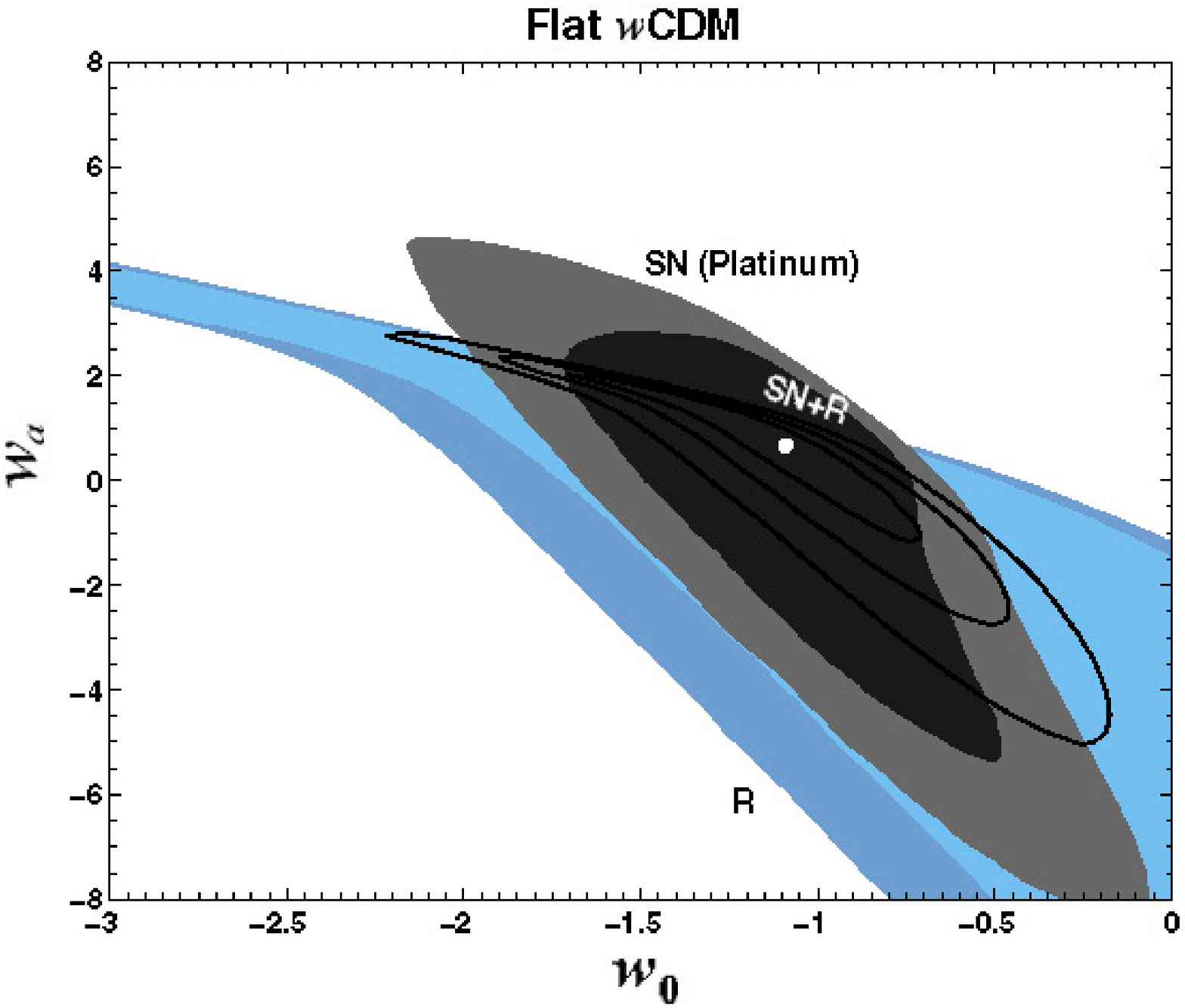}
\caption[]{Similar to Fig.~\ref{fig:wcdm2}, but with an equation of state $w(a)=w_0+w_a(1-a)$ for $0\le z \le 1089$. 
\label{fig:wcdm3}}
\end{center}
\end{figure}

\begin{figure}[hbp]
 \begin{center}
  \epsfysize=4truein
  \epsfxsize=5truein
    \epsffile{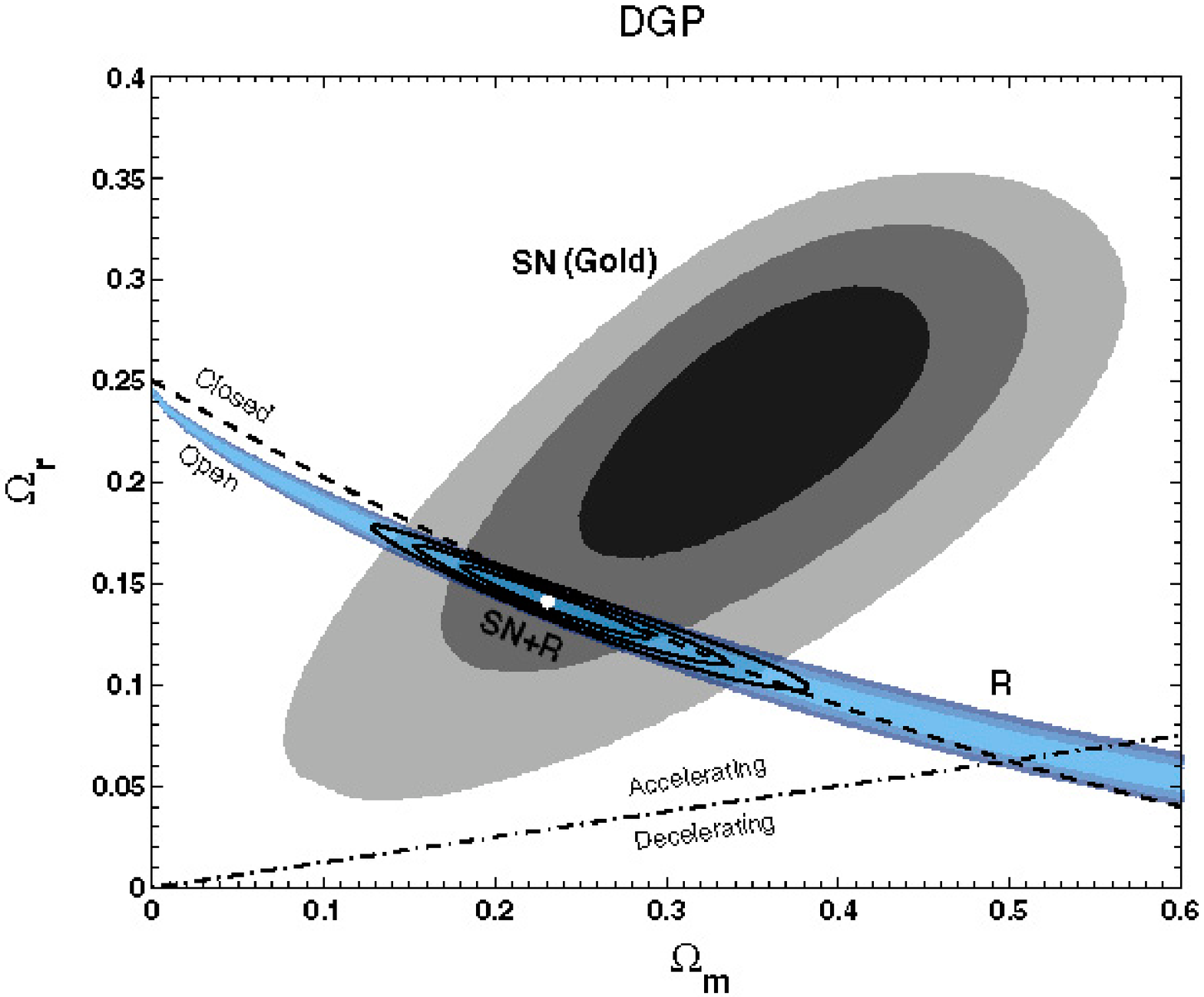},
  \epsfysize=4truein
  \epsfxsize=5truein
    \epsffile{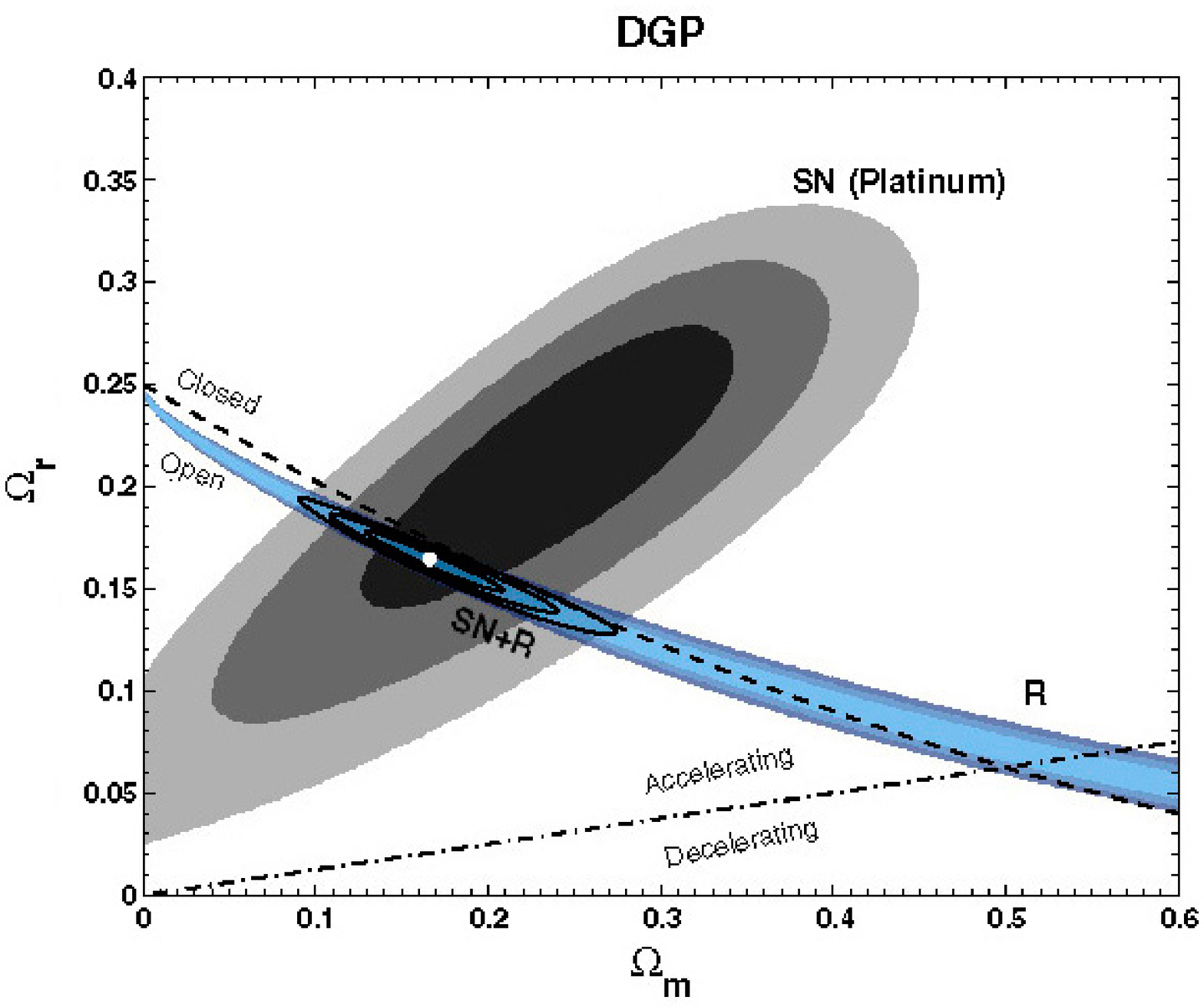}
\caption[]{Similar to Fig.~\ref{fig:lcdm}, but for the DGP model. 
\label{fig:dgp}}
\end{center}
\end{figure}

\begin{figure}[hbp]
 \begin{center}
  \epsfysize=4truein
  \epsfxsize=5truein
    \epsffile{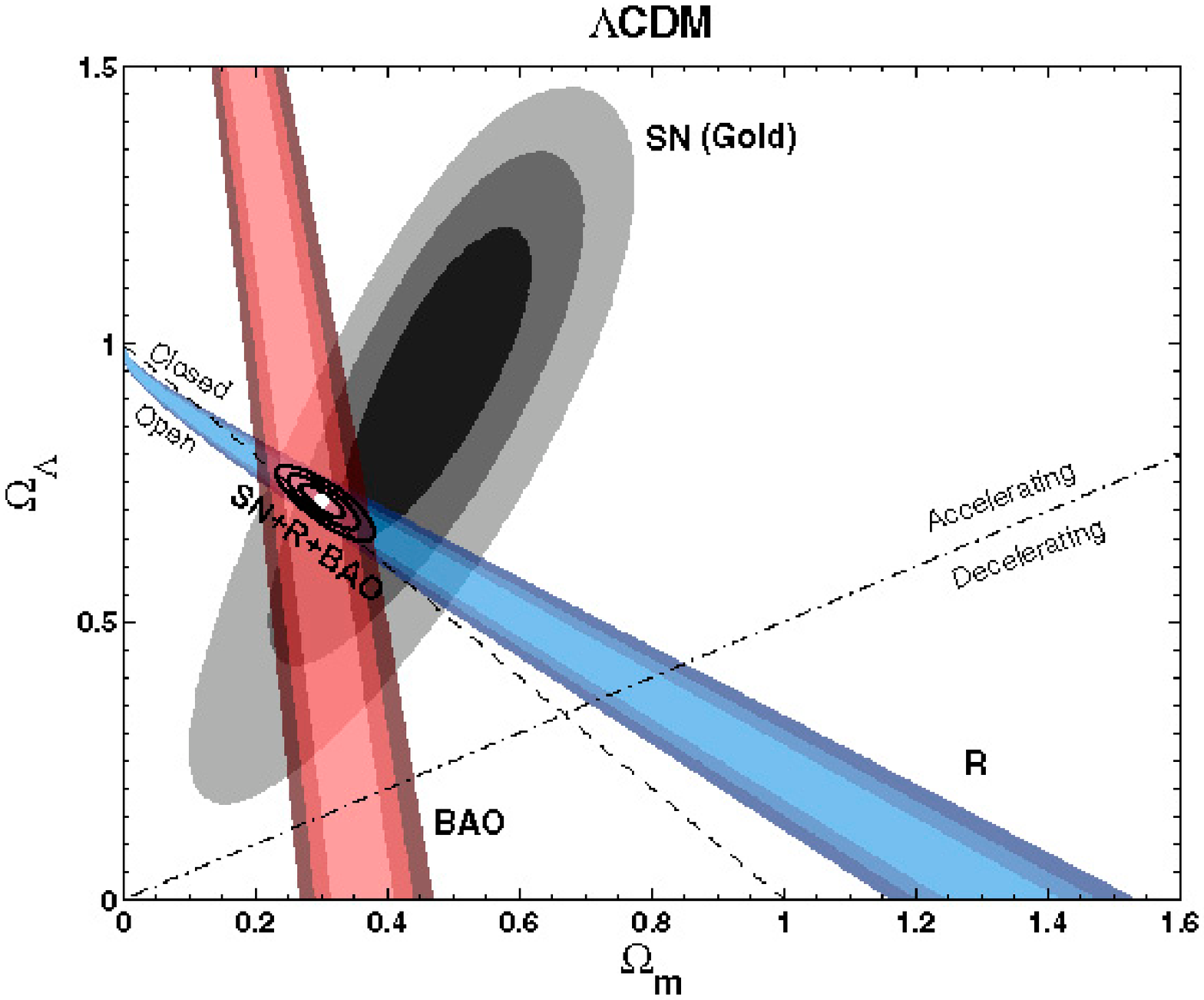},
  \epsfysize=4truein
  \epsfxsize=5truein
    \epsffile{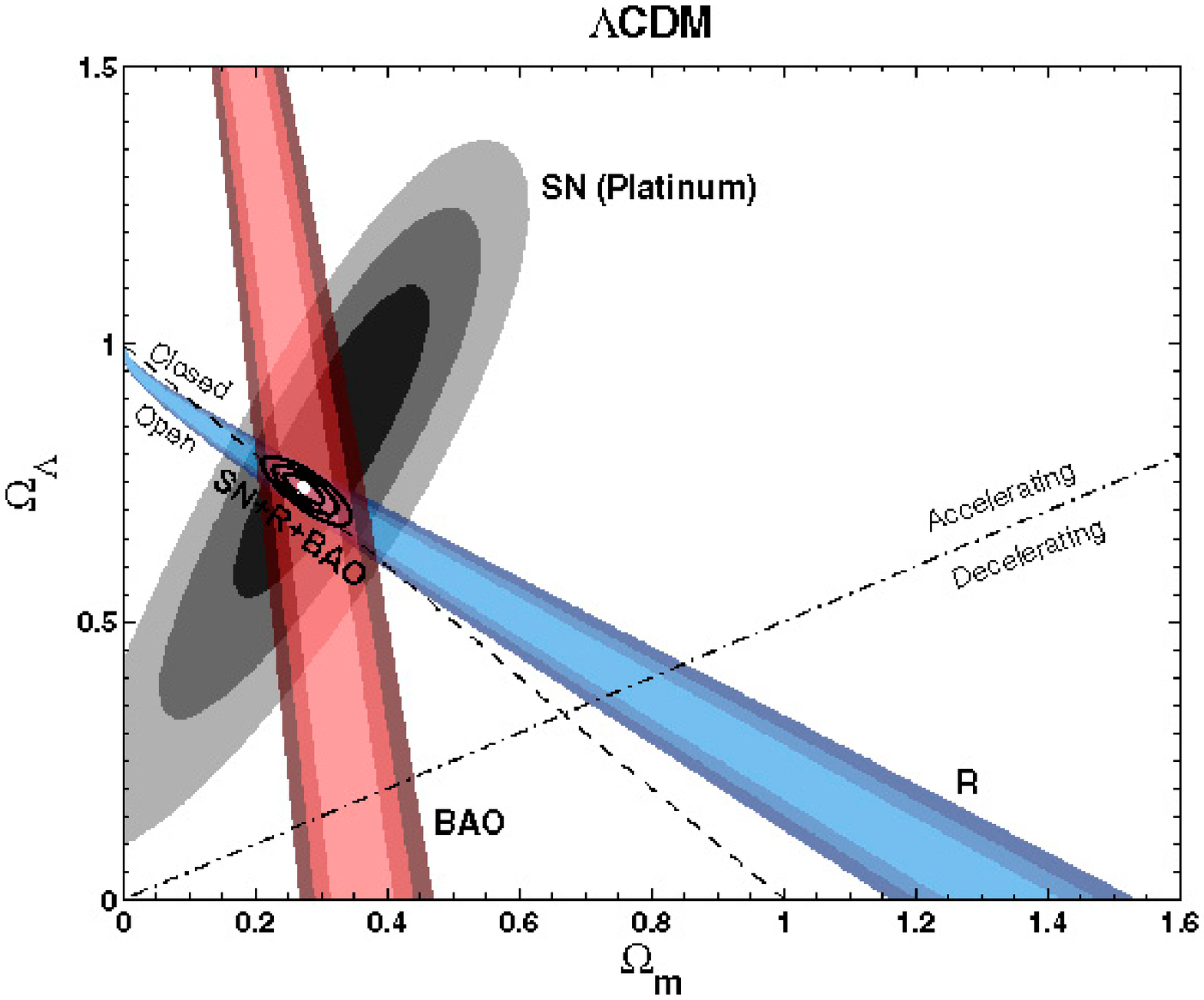}
\caption[]{Similar to Fig.~\ref{fig:lcdm}, but also showing the effect of the BAO constraint on the parameter space. 
\label{fig:bao1}}
\end{center}
\end{figure}

\begin{figure}[hbp]
 \begin{center}
  \epsfysize=4truein
  \epsfxsize=5truein
    \epsffile{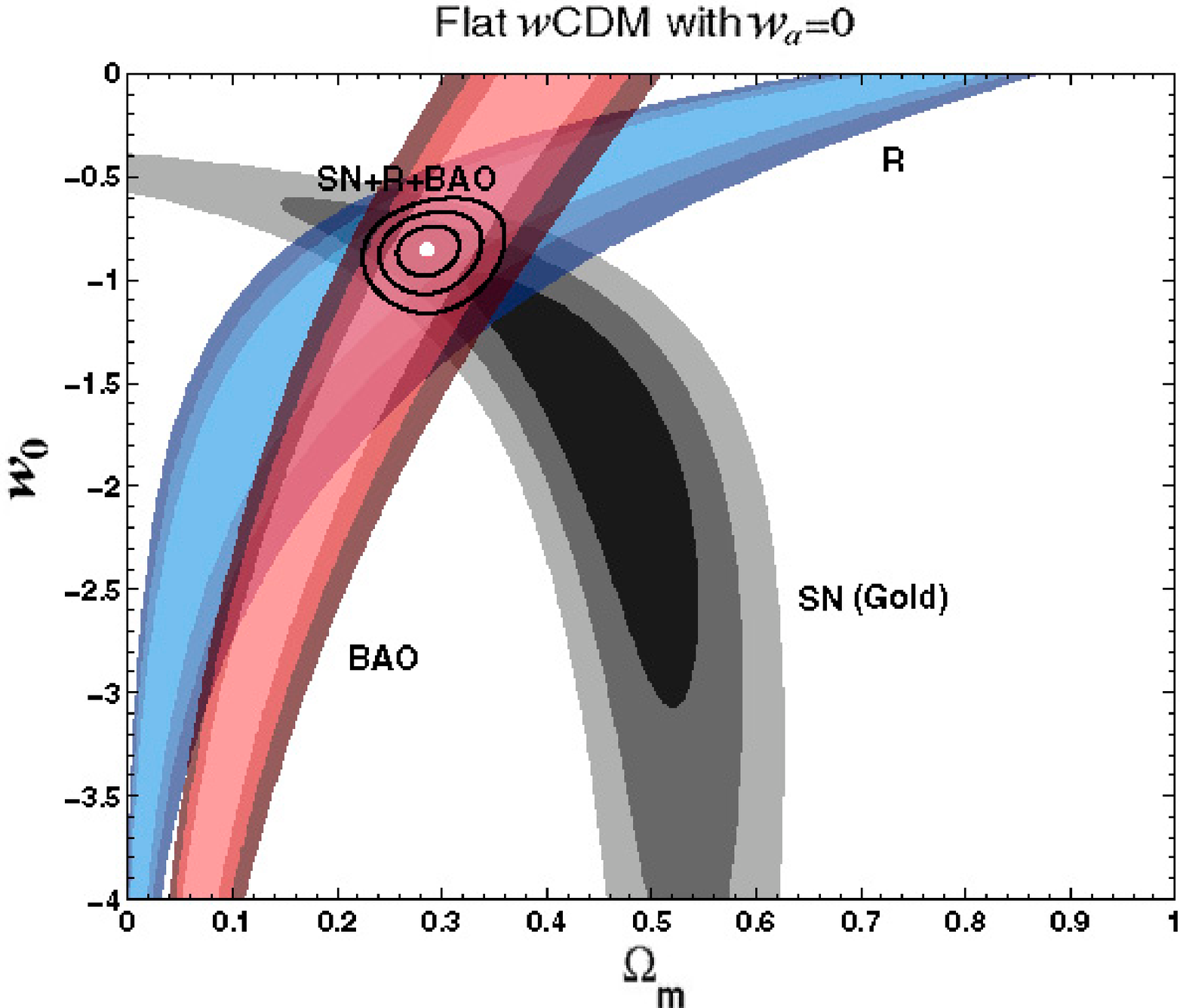},
  \epsfysize=4truein
  \epsfxsize=5truein
    \epsffile{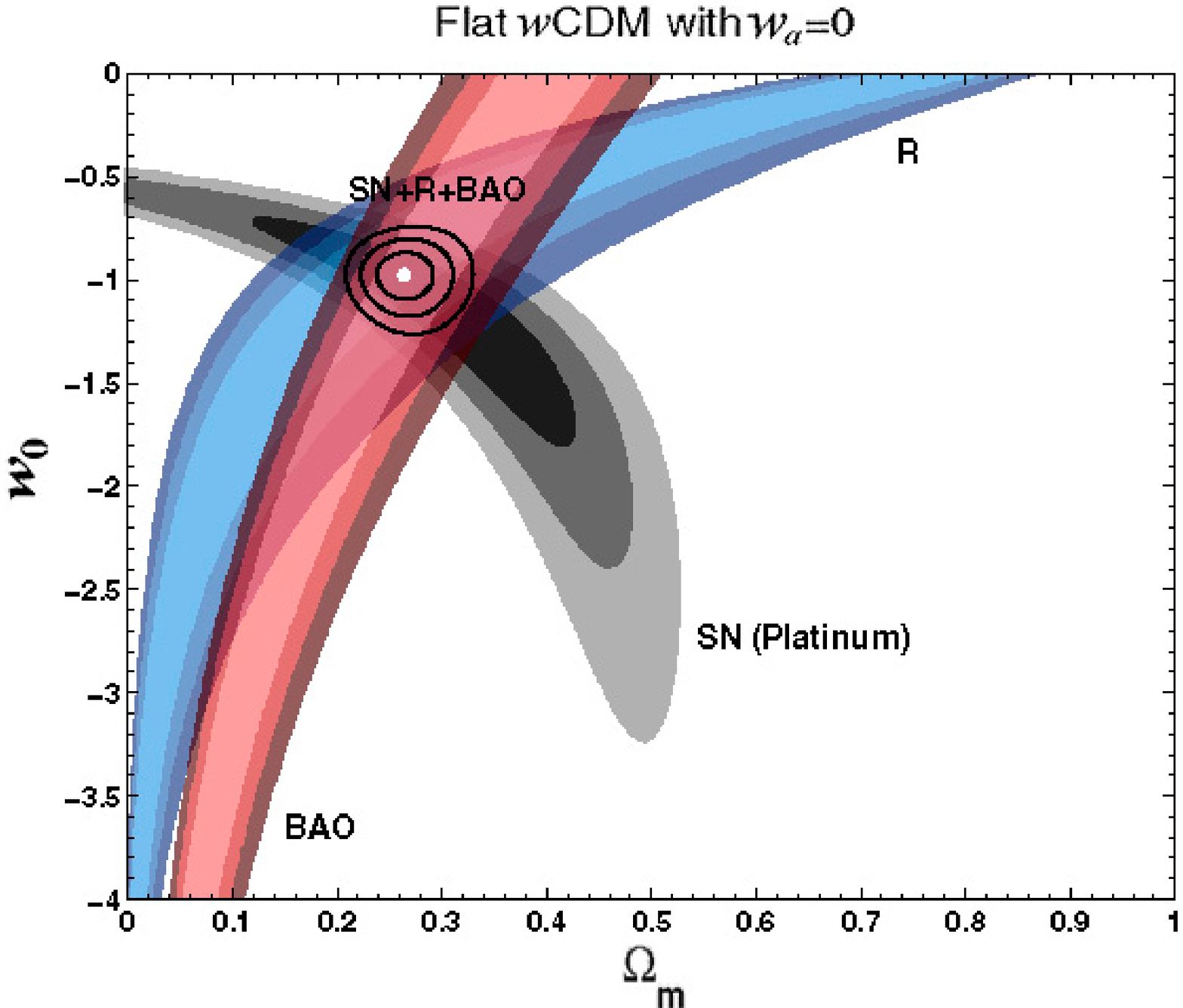}
\caption[]{Similar to Fig.~\ref{fig:wcdm1}, but also showing the effect of the BAO constraint on the parameter space.
\label{fig:bao2}}
\end{center}
\end{figure}

\end{document}